\documentclass[12pt]{elsarticle}

\usepackage{lineno,hyperref}
\usepackage{amssymb}
\usepackage{amsthm}
\usepackage{amsmath}

\usepackage{spverbatim}
\usepackage{bbold}
\usepackage{array}
\usepackage{caption}
\usepackage{etoolbox}
\AtBeginEnvironment{figure}{\addvspace{5mm}}
\usepackage{floatrow}
\usepackage{setspace}
\newtheorem{theorem}{Theorem}
\newtheorem{corollary}{Corollary}
\newtheorem{definition}{Definition}

\modulolinenumbers[5]

\journal{Journal of Mathematical Psychology}

\usepackage{natbib}

\begin{document}
\begin{frontmatter}

\title{A Better (Bayesian) Interval Estimate for Within-Subject Designs \tnoteref{t1}}
\tnotetext[t1]{This work was supported by discovery grants to Farouk Nathoo and Michael Masson from the Natural Sciences and Engineering Research Council of Canada. }

\author[label1]{Farouk S. Nathoo\corref{cor1}}
\ead{nathoo@uvic.ca}
\author[label2]{Robyn E. Kilshaw}
\author[label2]{Michael E. J. Masson\corref{cor1}}
\ead{mmasson@uvic.ca}

\cortext[cor1]{Corresponding authors}

\address[label1]{Department of Mathematics and Statistics, University of Victoria, Canada}
\address[label2]{Department of Psychology, University of Victoria, Canada}

\begin{abstract}
We develop a Bayesian highest-density interval (HDI) for use in within-subject designs. This credible interval is based on a standard noninformative prior and a modified posterior distribution that conditions on both the data \emph{and} point estimates of the subject-specific random effects. Conditioning on the estimated random effects removes between-subject variance and produces intervals that are the Bayesian analogue of the within-subject confidence interval proposed in Loftus and Masson (1994). We show that the latter interval can also be derived as a Bayesian within-subject HDI under a certain improper prior. We argue that the proposed new interval is superior to the original within-subject confidence interval, on the grounds of (a) it being based on a more sensible prior, (b) it having a clear and intuitively appealing interpretation, and (c) because its length is always smaller. A generalization of the new interval that can be applied to heteroscedastic data is also derived, and we show that the resulting interval is numerically equivalent to the standardization method discussed in Franz and Loftus (2012); however, our work provides a Bayesian formulation for the standardization method, and in doing so we identify the associated prior distribution.

\end{abstract}


\end{frontmatter}


\doublespace
\section{Introduction}
Loftus and Masson (1994) noted that in a within-subject design, the standard confidence interval for the mean of the response at a particular level of the independent variable, may in fact not be practical for interpretation. This is because, at different levels of the independent variable, such confidence intervals can show substantial overlap, resulting in a genuine trend of within-subject effects being hidden---especially when the between-subject variability is relatively high. Consequently, even if the corresponding repeated-measures (within-subject) analysis of variance (ANOVA) indicates a highly significant $F$ statistic, it may not be possible to make any strong inference about the ordering of the condition means based on standard confidence intervals. Conflicting evidence drawn from the standard confidence interval (CI) and the repeated-measures ANOVA arises because the between-subject variance, which is irrelevant in the repeated-measures ANOVA, partially determines the length of a standard CI.

Given the irrelevance of the between-subject variance in a repeated-measures ANOVA, Loftus and Masson (1994) proposed the within-subject CI as an interval estimate for use in within-subject designs, where the length of the interval does not depend on the amount of between-subject variability. The within-subject CI is based on a transformation of the data that removes this source of variability. The resulting interval is smaller in length (assuming cross-measurement correlation is positive) than a standard CI; however, a significant downside is that at nominal level $1-\alpha$, the coverage probability of the within-subject CI can be far less than $1-\alpha$. As a result, the within-subject CI is not a valid $1-\alpha$ CI in the usual sense. Nevertheless, it can be used as a graphical tool that may uncover the true pattern of within-subject effects (that could otherwise be hidden within a set of standard CIs), while at the same time depicting the variability that is of scientific interest in within-subject designs.

A more significant challenge to the interpretation of CIs, which can be extended to the special case of within-subject intervals, is that even experienced researchers can hold erroneous views of what CIs actually mean. This has been discussed by, among others, Hoekstra, Morey, Rouder, and Wagenmakers (2014). These misunderstandings include the idea that a CI has a specified probability $1-\alpha$ of containing the true value of the parameter. Interpretations such as this are perpetuated by text books and other sources (e.g., Cumming 2014; Masson \& Loftus, 2003) that do not accurately reflect the concept of CIs as originally proposed by Neyman (1937). Under Neyman's definition, a CI is an interval generated by a procedure that, with repeated sampling, has a specified probability of containing the true value of the parameter. It is important to realize that this is a pre-data specification of confidence. A CI realized from a particular sample of data, however, cannot itself be linked back to the degree of confidence that was specified before the sample was collected (Morey, Hoekstra, Rouder, Lee, \& Wagenmakers, 2016a; Morey, Hoekstra, Rouder, \& Wagenmakers, 2016b).

The common misunderstanding of how to interpret CIs is associated with a tendency to treat them as credible intervals---ranges of values that are a posteriori likely to include the parameter of interest. Given the current advocacy of the use of CIs in psychological science (e.g., Cumming, 2013, 2014) and these concerns about widespread misinterpretation of CIs, there may be great value in adopting Bayesian credible intervals as the standard means of expressing estimates of parameter values and the relative precision of those estimates (Kruschke, 2013; Morey et al., 2016a).

We note, however, that there is an important area where CIs have been widely applied, but no method of computing a meaningful Bayesian credible interval has yet been developed. Namely, for repeated-measures or within-subject designs, Loftus and Masson (1994) introduced a widely accepted method (now cited over 2200 times according to Google Scholar) for computing within-subject CIs that reflect the relative magnitudes of sample means. In order to avoid the aforementioned problems associated with interpreting CIs, however, we describe in this paper our development of a Bayesian analogue of this within-subject CI.  It is in fact surprising that in the 24 years since the publication of the Loftus and Masson article, no method for Bayesian within-subject interval estimation has been developed, as far as we are aware.  Nevertheless, such a development is certainly well-motivated, if only because the move toward a posterior probability interpretation of the within-subject interval will be intuitively more appealing to most scientists. 

We therefore present a Bayesian highest density interval (HDI), where the length of this HDI does not depend on the amount of between-subject variability in the data. In addition, we go one step further and properly characterize the posterior probability of such an interval.  This latter aspect of our work is a particularly useful new development, in that it allows users of the within-subject interval to understand with greater clarity the sense in which such an interval is a valid $1-\alpha$ interval estimate.  As with its classical counterpart, the within-subject HDI is not a standard HDI in the usual sense; nevertheless, our construction will give the user a concrete and more direct understanding of its associated posterior probability at nominal level $1-\alpha$.

To develop the Bayesian within-subject HDI, we adopt a rather non-standard approach and base inference on a \emph{modified posterior distribution} that conditions not only on the data, but also on the subject-specific random effects in the corresponding mixed effects model.  We note that using a modified posterior distribution (as opposed to the standard posterior distribution for Bayesian inference) may be viewed as somewhat controversial, and as a departure from the standard Bayesian paradigm. We call our new paradigm conditional Bayesian inference.

Another example where a modified posterior distribution is used for statistical inference is mean-field variational Bayes inference (e.g., Ormerod \& Wand, 2010; Ostwald et al., 2014; Nathoo et al., 2014)---a method in which the modified posterior distribution is assumed to take a factorized form, and the modification is made to facilitate faster computation. In our case, the modification to the posterior distribution is used to eliminate the contribution of between-subject variability, and a key point here is that this variability is not of scientific interest in within-subject designs. Removing this component of variability was the motivation for the original within-subject CI (Loftus \& Masson, 1994), and this approach is now an industry standard for use in within-subject designs in psychology.

As our approach to constructing the within-subject HDI is conditional on random effects that are not known, it is necessary to estimate the random effects. We estimate these effects using maximum likelihood and then condition on the estimated random effects. Thus, our use of plug-in estimates gives our approach an empirical Bayes flavor, although it is not our objective to estimate the prior from the data as it is with empirical Bayes. The idea of using conditional inference and empirical Bayes to improve efficiency in the presence of many nuisance parameters is considered by Liang and Tsou (1992), and conditional inference in the presence of nuisance parameters is discussed by Cox and Reid (1987). 

Using the idea of a modified posterior distribution, we are able to derive the original within-subject CI as a within-subject Bayesian HDI corresponding to a particular improper prior. To be clear, we are not advocating treating CIs as credible intervals in general, nor are we advocating treating frequentist and Bayesian intervals as interchangeable. The problems associated with this have been discussed in Hoekstra et al. (2014) and elsewhere. The numerical equivalence between the within-subject CIs and the within-subject HDIs presented in this paper is useful in that it allows users of \emph{these specific methods} to apply a posterior probability interpretation---an interpretation which is undoubtedly more appealing to most practitioners. 

In addition, the Bayesian formulation clarifies for the user what the underlying prior distribution actually is and, as we shall demonstrate, that the standard within-subject CI, when formulated as a Bayesian within-subject HDI, is based on a prior distribution that may be questionable. We therefore develop a new Bayesian within-subject HDI based on a standard noninformative prior distribution, and we show that this new interval \emph{always has shorter length than the original within-subject CI of Loftus and Masson (1994)}.

The noninformative prior that we use to develop the proposed within-subject HDI is a standard noninformative prior commonly used for normal models, and was also considered in Rouder, Morey, Speckman, and Province (2012). It is well known that the use of noninformative priors can cause problems for Bayesian hypothesis testing with the Bayes factor, when these priors are assigned to parameters that are not common to the models being compared (e.g., Wetzels, Grasman, \& Wagenmakers, 2012). In our context, however, the use of a noninformative prior causes no theoretical difficulty. Although the prior is improper, we show that the corresponding posterior distribution is always proper and can thus be used to construct an interval estimate with the specified posterior probability.

One can argue against our proposed approach of conditioning on estimated random effects as providing '\emph{false certainty}' in its double use of the data; first to estimate the between-subject variability and then again to construct the interval estimate, given the estimated between-subject variability. Nevertheless, this same criticism can also be launched against the original non-Bayesian within-subject CI, despite this approach's extensive use and acceptance in psychology for the analysis of repeated-measures designs. As mentioned earlier, the practical justification for the within-subject CI, and perhaps the reason for its continued wide use, is that the within-subject interval removes the component of variability that \emph{is not} of scientific interest in this experimental design. For the reasons discussed above, this interval addresses the limitations of the standard CI (i.e., between-subject variance masking the true pattern of within-subject effects), while providing the researcher with a depiction of the variability that \emph{is} of relevance in within-subject designs.

With regard to the conditional Bayesian approach we have adopted and the alternative of a fully Bayesian approach,  we acknowledge that our use of plug-in estimates of the random effects implies that the uncertainty in estimating the subject-specific random effects is not propagated to the width of the HDI. Estimating the random effects is necessary because the intervals are based on a conditional posterior distribution, where we are conditioning on unknown quantities, the subject-specific random effects, in order to remove the uninteresting component of variability. We show that this approach, under a certain improper prior, leads to the original interval proposed by Loftus and Masson (1994) which demonstrates that it is a reasonable approach to removing between-subject variance. A fully Bayesian approach precludes conditioning on random effects, and thus would not yield within-subject Bayesian inference. It is the conditioning on estimated random effects that removes the uncertainty that is not of interest, and this conditioning precludes fully Bayesian inference. 

The resulting Bayesian HDI leads to two main advantages over the classical within-subject CI. First, it is worth reiterating that the move towards a posterior probability interpretation of the within-subject interval will be intuitively more appealing to most scientists. Second, our use of the modified posterior distribution allows us to attach a direct modified posterior probability to the within-subject HDI; a modification which grants the user a precise understanding of what the associated posterior probability actually is when the within-subject HDI is constructed at nominal level $1-\alpha$.

We will develop the Bayesian within-subject interval for two cases, each corresponding to an underlying mixed effects model. In the first case, we will assume that the error variance of the response, $\sigma_{\epsilon}^{2}$, is constant across different levels of the experimental factor. In the second case, we relax this assumption and develop a within-subject HDI that can be applied to heteroscedastic data.

The remainder of the paper proceeds as follows. In Section 2, we formulate the new Bayesian within-subject HDI and discuss its connection to the original within-subject CI. In Section 3, we present a Bayesian within-subject HDI that can be applied to heteroscedastic data. Section 4 presents two practical examples, with a tutorial aspect designed to help researchers who are new to Bayesian methods become familiar with applying these methods to their data. In these examples, we make comparisons between the classical within-subject CI and the new Bayesian within-subject HDI, and we also compare these intervals to the classical between-subject interval estimate. Section 5 concludes the paper with a discussion.

\section{Formulation of the Bayesian HDI for Within-Subject Designs}
Consider a single-factor repeated-measures design with the corresponding mixed effects model
\begin{equation}
\label{mixed_model}
  Y_{ij}=\mu_j+b_i+\epsilon_{ij},\;\;\,\;\;\epsilon_{ij} \stackrel{iid}\sim\text{N}(0,\,\sigma_{\epsilon}^{2}),\\
   \;\;i = 1,\dots\,,N;\;\;j = 1,\dots\,,C,
\end{equation}
where $Y_{ij}$ represents the response obtained from the $i^{th}$ subject under the $j^{th}$ level of the experimental manipulation; $\mu_{j}$ is the mean of the response at the $j^{th}$ level; $N$ is the number of subjects; $C$ is the number of levels; and $b_{i}$ is a mean-zero random effect for the $i^{th}$ subject.

Loftus and Masson (1994) proposed the within-subject confidence interval (CI) for the condition mean $\mu_j$ based on the idea of first transforming the data to remove the between-subject variability. The resulting interval has smaller length than a standard confidence interval for $\mu_{j}$, and its construction is motivated by the following notion: since between-subject variance typically plays no role in the statistical analyses of within-subject designs, it can legitimately be ignored. Hence, an appropriate confidence interval can be based on the standard within-subject error term.

The $100(1-\alpha)\%$ within-subject CI takes the form
\begin{equation}
\label{LM}
    M_{. j}\pm\sqrt{\frac{SS_{S\times C}}{N(N-1)(C-1)}}\;\;[\textup{criterion}\ t_{(C-1)(N-1)}]_{\frac{\alpha}{2}}
\end{equation}
where $M_{. j} = \frac{1}{N}\sum_{i=1}^{N}Y_{ij}$ is the estimated condition mean;  
$$
SS_{S\times C} = \sum_{j=1}^C\sum_{i=1}^NY_{ij}^2 - C\sum_{i=1}^NM_{i .}^2 - N\sum_{j=1}
   ^CM_{. j}^2 + CNM^2
$$
is the interaction sum-of-squares, with $M_{i .} =  \frac{1}{C}\sum_{j=1}^{C}Y_{ij}$ being the mean of the data obtained from the $i^{th}$ subject (note then that $M_{i .}$ and $M_{. j}$ denote the subject and condition means respectively); and $M = \frac{1}{N C}\sum_{i=1}^{N}\sum_{j=1}^{C}Y_{ij}$ is the overall mean.

The Loftus and Masson (1994) within-subject confidence interval is identical to that proposed in Cousineau (2005) and Morey (2008) when the latter use $(C-1)(N-1)$ degrees-of-freedom, except that the former interval is based on a pooled standard deviation and the latter uses un-pooled estimates. Loftus and Masson (1994) note that the interval estimate (\ref{LM}) is \emph{not a bona fide confidence interval}, since it is based only on interaction variance and is not a function of the between-subject variability. Although removing the latter component of variability is in fact the motivation for the within-subject interval, an important consequence is that the within-subject interval will not have a coverage probability equal to the nominal coverage probability of $1-\alpha$. Further, it is not clear what the (frequentist) coverage probability of such an interval actually is. The authors justify its validity in a typical within-subject design, by noting that the within-subject CI at level $1-\alpha$ has the property that its length is related by a factor of $\sqrt{2}$ to the standard $1-\alpha$ confidence interval around the difference between two condition means (assuming that the variance is constant across conditions). The latter point provides the user with some (albeit indirect) notion in which (\ref{LM}) is a $1-\alpha$ interval estimate, and indeed, the methodology is now in common use for the analysis of repeated-measures designs in psychology.

As an alternative derivation of a within-subject credible interval, we adopt a Bayesian approach that leads to a within-subject interval that has a more direct interpretation in terms of modified posterior probability. As with the original within-subject CI, our goal is to develop an interval estimate that has, in some sense, removed the between-subject variance such that a more efficient interval is obtained. Rather than transforming the data, we do this by using a modified posterior distribution that conditions both on the data, and on point estimates of the subject-specific random effects. This modified posterior is constructed to improve the efficiency of the interval estimate for the parameter of interest $\{\mu_{j}\}$, in the presence of what are essentially many nuisance parameters $\{b_{i}\}$. The latter are necessary to characterize the differences among the subjects, but are typically of no scientific interest in within-subject designs. In our context, the nuisance parameters are eliminated through conditioning, but since the $b_{i}$ are unknown, they are replaced by estimates $\hat{b}_{i}$. We use the maximum likelihood estimate $\hat{b}_{i} = M_{i .} - M$ which is obtained by solving $\partial \log f(\boldsymbol{Y};\boldsymbol{\mu},\sigma^{2}_{\epsilon},b_{1},\dots,b_{n})/\partial b_{i} = 0$ and $\partial \log f(\boldsymbol{Y};\boldsymbol{\mu},\sigma^{2}_{\epsilon},b_{1},\dots,b_{n})/\partial \boldsymbol{\mu} = \mathbf{0}$; where $\boldsymbol{Y}$ denotes the data, $\boldsymbol{\mu} = (\mu_{1},\dots,\mu_{C})^{T}$, and $f(\boldsymbol{Y};\boldsymbol{\mu},\sigma^{2}_{\epsilon},b_{1},\dots,b_{n})$ is the conditional density of $\boldsymbol{Y}$ given $\boldsymbol{\mu}$, $b_{1},\dots,b_{n}$, and $\sigma^{2}_{\epsilon}$. The usual specification of the normal or other parametric distribution for the random effects is not needed as we do not integrate over the random effects. The modified posterior distribution is a conditional distribution where we condition on both the data and the random effects. Thus, no distribution for the random effects is assumed or required when computing the interval, and in this sense our approach can be considered semiparametric.

One may understand this idea further by considering the simple inequality taught in introductory probability, with its presentation modified for our context as $$Var[\mu_{j}|\boldsymbol{Y}] \ge E[Var[\mu_{j}|\boldsymbol{Y}, b_{1},\dots,b_{n}]],$$ 
which indicates, in expectation, the conditional posterior variance of the parameter given the subject-specific random effects, is less than or equal to the unconditional posterior variance of this parameter. We thus use this additional conditioning on subject-specific random effects as a mechanism for constraining the variability so that the component of posterior variability that is not of scientific interest is removed. Importantly, we do this while at the same time obtaining an interval whose posterior probability, albeit conditional on the estimated random effects, can be quantified exactly as $1-\alpha$. This is an improvement over the original notion of the within-subject CI which at level $1-\alpha$ \emph{does not} have a (frequentist) coverage probability of $1-\alpha$; furthermore, its actual coverage probability is not specified nor clear.

Thus, although a standard Bayesian HDI is based on the posterior density $p(\mu_{j} | \boldsymbol{Y})$, our proposal for a Bayesian within-subject HDI is based on the modified posterior density $p(\mu_{j} | \boldsymbol{Y},\hat{b}_1,\dots\,,\hat{b}_N)$, where $\hat{b}_i = M_{i .} - M$ is a point estimate of $b_i$, so that our approach treats estimated random effects as known and fixed values that are estimated with maximum likelihood.

\begin{definition}
For the mixed effects model (\ref{mixed_model}), the $100(1-\alpha)\%$ Bayesian within-subject HDI for $\mu_{j}$ is the set $\boldsymbol{I} \subset \mathbb{R}$ with
$$
\boldsymbol{I} = \{\mu_{j}: p(\mu_{j} | \boldsymbol{Y},\hat{b}_1,\dots\,,\hat{b}_N) \ge k\}
$$
with  $k$ chosen as the largest number so that
\begin{equation}
\label{posterior_prob}
Pr(\mu_{j} \in \boldsymbol{I} |\boldsymbol{Y}, \hat{b}_1,\dots\,,\hat{b}_N) = \int_{\boldsymbol{I}} p(\mu_{j} | \boldsymbol{Y},\hat{b}_1,\dots\,,\hat{b}_N) d \mu_{j} = 1-\alpha.
\end{equation}
\end{definition}
Definition 1 gives us a precise notion, in terms of modified posterior probability, of how to construct and subsequently interpret the within-subject HDI. It is worth mentioning that although the within-subject HDI is defined in terms of its modified posterior probability, nothing prevents the practitioner from also calculating its associated \emph{unconditional posterior probability} $Pr(\mu_{j} \in \boldsymbol{I} |\boldsymbol{Y})$, and such a calculation is straightforward given the posterior distribution or a Monte Carlo representation of this distribution. 

\begin{theorem}
Assuming the noninformative prior $\pi(\mu_{1}, \dots, \mu_{C},\sigma_{\epsilon}^{2})\propto\frac{1}{\sigma_{\epsilon}^{2}}$ and the mixed effects model (\ref{mixed_model}), the posterior distribution conditioning also on the point estimates $\hat{b}_i = M_{i .} - M$ takes the form
$$
\mu_{j} | \boldsymbol{Y}, \hat{b}_1,\dots\,,\hat{b}_N \sim t_{C(N-1)}\bigl(M_{. j},\frac{SS_{S\times C}}
   {N(N-1)C}\bigr)
$$
\end{theorem}
\noindent The proof is given in the Appendix.
The Jeffreys prior adopted here\\ $\pi(\mu_{1}, \dots, \mu_{C},\sigma_{\epsilon}^{2}) \propto\frac{1}{\sigma_{\epsilon}^{2}}$ is a noninformative prior and has the advantage of being invariant to one-to-one transformations.

\begin{corollary}
Under the conditions of Theorem 1, the $100(1-\alpha)\%$ within-subject Bayesian HDI satisfying equation (\ref{posterior_prob}) takes the form
\begin{equation}
\label{new_HDI}
    M_{. j}\pm\sqrt{\frac{SS_{S\times C}}{N(N-1)C}}\;\;[\textup{criterion}\ t_{C(N-1)}]_{\frac{\alpha}{2}}
\end{equation}
\end{corollary}
\noindent The proof is given in the Appendix. 

We notice that this Bayesian within-subject HDI has a simple form that is very similar to the original within-subject CI (\ref{LM}). Furthermore, the original within-subject CI can also be interpreted as a Bayesian within-subject HDI satisfying equation (\ref{posterior_prob}) under a specific improper prior. In addition, the proposal (\ref{new_HDI}) is equivalent to that considered in Cousineau (2005) and Morey (2008) when both use $C(N - 1)$ degrees-of-freedom, with the only difference being that the latter intervals do not use a pooled estimate of the standard deviation. A referee has noted that there can be a tiny advantage to using separate estimates in some settings as this corresponds to a more flexible model. In this case the intervals considered in Cousineau (2005) and Morey (2008) may have shorter length. In addition, Morey (2008) proved for the hierarchical model assumed here that using separate estimates of the standard deviation has advantages with respect to bias which is then reflected in the corresponding intervals. Of course, this will come at the cost of increased variance as the number of parameters to be estimated increases.

\begin{theorem}
Assuming the improper prior $\pi(\boldsymbol{\mu}, \sigma_{\epsilon}^2)\propto\Bigl[\frac{1}{\sigma_{\epsilon}^2}\Bigr] ^{\frac{-N+3}{2}}$
and the mixed effects model (\ref{mixed_model}), the  $100(1-\alpha)\%$ within-subject Bayesian HDI for $\mu_{j}$ is precisely the original within-subject CI of Loftus and Masson (1994) given in equation (\ref{LM}) .
\end{theorem}
\noindent The proof is given in the Appendix.

Thus, from Theorem 2, we are able to interpret the original within-subject CI for $\mu_{j}$ as a Bayesian within-subject interval with modified posterior probability $1-\alpha$, as specified in equation (\ref{posterior_prob}). This is, in and of itself, an interesting new interpretation of the original within-subject CI that provides substantial new clarity on how to interpret its associated posterior probability at nominal level $1-\alpha$. Nevertheless, our proposed new within-subject interval (\ref{new_HDI}) based on the noninformative prior seems to be a better within-subject interval for at least two reasons:

\begin{enumerate}
\item First, the new interval is \emph{always shorter} than the original within-subject interval, \emph{even though both have the same modified posterior probability of $1-\alpha$}. More specifically, letting $L_{NKM}$ denote the length of the new interval, $L_{LM}$ denote the length of the original interval, and $R = L_{NKM}/L_{LM}$, we have that
\begin{equation}
\label{length_ratio}
   R = \frac{L_{NKM}}{L_{LM}}=\underbrace{\sqrt{\frac{C-1}{C}}}_{<1}\underbrace{\frac{[\textup{criterion}\ 
   t_{C(N-1)}]_{\frac{\alpha}{2}}}{[\textup{criterion}\ t_{(N-1)(C-1)}]_{\frac{\alpha}{2}}}}_{<1} <1.
\end{equation}
The comparison of the lengths of these two intervals is appropriate and meaningful since: (1) both are $1-\alpha$ within-subject HDI's so that we are comparing two Bayesian within-subject intervals under different priors; (2) both are centred at the same point, namely, $M_{. j}$; so it is relevant to point out that the proposed within-subject HDI has smaller length than the original within-subject CI. 
\item Second, the new interval is based on a standard noninformative prior distribution, $\pi(\mu_{1}, \dots, \mu_{C},\sigma_{\epsilon}^{2})\propto\frac{1}{\sigma_{\epsilon}^{2}}$, that is far more reasonable than the prior distribution corresponding to the original interval $\pi(\boldsymbol{\mu}, \sigma_{\epsilon}^2)\propto\Bigl[\frac{1}{\sigma_{\epsilon}^2}\Bigr] ^{\frac{-N+3}{2}}$. Although both priors are improper, and both lead to a proper posterior distribution, the former prior is a decreasing function of $\sigma_{\epsilon}^{2}$, whereas the latter is an increasing and unbounded function of $\sigma_{\epsilon}^{2}$ (assuming $N>4$). Giving larger prior weight to larger values of $\sigma_{\epsilon}^{2}$ without bound seems an unreasonable prior; \emph{thus, practitioners should be aware that the widely used within-subject CI corresponds to an apparently unreasonable prior.}
\end{enumerate}
We therefore claim that the new interval (\ref{new_HDI}) is a better within-subject interval. This claim also corresponds to a claim that the Jeffreys prior leads to a shorter interval than the prior associated with Loftus and Masson (1994), while both intervals are centred at the same point estimate. We note that this does not necessarily imply that the new interval is optimal in any sense. In addition, it is certainly possible to use a prior that will result in an even smaller interval than Jeffreys prior used here (consider for an extreme example, a point mass prior). However, this does not imply that such an interval is necessarily better since its centre might be a biased estimator of location and the bias can be the result of a strong prior. In the context of our comparison of  (\ref{new_HDI})  with the within-subject CI of Loftus and Masson (1994), this point about bias does not apply since these intervals are centred at the same point and both are symmetric.

\section{Dealing with Heteroscedastic Data}

A key assumption of the mixed effects model (\ref{mixed_model}) is that the error variance $\sigma_{\epsilon}^2$ is constant across the different levels of the experimental manipulation---an assumption that can be violated in psychological experiments. One possible solution is to apply a transformation to the response that makes the variance of the transformed response stable. Another solution is to simply expand the model so that it allows for this behavior. We adopt the second solution here and derive a Bayesian within-subject HDI based on the more general one-way mixed effects model
\begin{equation}
\label{mixed_H}
   Y_{ij}=\mu_j+b_i+\epsilon_{ij}\;\;\,,\;\;\epsilon_{ij} \stackrel{ind}\sim\text{N}(0,\,\sigma_{j}^{2})\\
   \;\;i = 1,\dots\,,N;\;\;j = 1,\dots\,,C,
\end{equation}
\noindent where now Var[$\epsilon_{ij}] = \sigma_j^2$ depends on the level $j$ of the experimental condition. As before, the within-subject Bayesian HDI is defined based on the conditional posterior density function $p(\mu_{j}\mid\boldsymbol{Y},\hat{b}_1,\dots\,,\hat{b}_N)$, where $\hat{b}_i = M_{i .}-M$. 

\begin{theorem}
Assuming the prior
\begin{equation*}
   \pi(\boldsymbol{\mu}, \sigma_1^2,\dots,\sigma_C^2)\propto\prod_{j=1}^C\sigma_j^{-2},
\end{equation*}
and the heteroscedastic mixed effects model (\ref{mixed_H}), the  $100(1-\alpha)\%$ within-subject Bayesian HDI for $\mu_{j}$ satisfying 
$
Pr(L_{j} \le \mu_{j} \le U_{j}|\boldsymbol{Y}, \hat{b}_1,\dots\,,\hat{b}_N) = 1-\alpha
$,
takes the form
\begin{equation}
\label{interval_HV}
  M_{. j}\pm SEM_{j}^{Norm}[\text{criterion}\ t_{N-1}]_{\frac{\alpha}{2}}
\end{equation}
\begin{gather*}
   \text{where}\;\;SEM_{j}^{Norm}=\sqrt{\frac{1}{N(N-1)}\sum_{i=1}^N(Y_{ij}^{'}-M_{. j})^2}\\
   Y_{ij}^{'}=Y_{ij}-M_{i .}+M.\\
\end{gather*}
\end{theorem}
\noindent The proof is provided in the Appendix. 

We note that this interval is precisely the standardization method discussed in Franz and Loftus (2012) and elsewhere, but we have provided here a Bayesian justification for this method along with a precise interpretation (in terms of modified posterior probability) of the resulting interval estimate. The interval (\ref{interval_HV}) also shows rigorously that the intervals proposed in Cousineau (2005) and Morey (2008) with $N -1$ degrees-of-freedom are an adequate solution when homogeneity of variances is violated in the data. The method is extremely simple to apply, and its implementation merely requires a standardization of the data $ Y_{ij}^{'}=Y_{ij}-M_{i .}+M$, after which the usual interval estimate used in between-subjects designs is constructed. We note again, however, that our contribution is the derivation of this method as a Bayesian within-subject HDI which allows for a completely novel interpretation and justification, based on its modified posterior probability (\ref{posterior_prob}). 

Standardization methods have also been discussed by Cousineau (2005) who proposed a simple alternative to the Loftus and Masson CIs that does not assume sphericity. This approach also removes individual differences in the data through a transformation. This same procedure was also described by Loftus and Masson (1994) to illustrate the process of removing individual differences from data rather than for computing the CI. Standardization methods are also discussed in Morey (2008) and Baguley (2012). Morey (2008) pointed out that Cousineau's (2005) approach produces intervals that are consistently too narrow because the standardization procedure induces a positive covariance between standardized scores within a condition, introducing bias into the estimates of the sample variances. Morey (2008) suggests a simple correction to the Cousineau (2005) approach, in which the half-width of the CI is rescaled by a factor of $\sqrt{\frac{C}{C-1}}$. The presence of this correction factor is now commonly considered in the Cousineau (2005) and Morey (2008) method and is unambiguously present in all subsequent publications (O'Brien \& Cousineau, 2014;  Baguley, 2012;  Cousineau \& O'Brien, 2014).

Related to this, Franz and Loftus (2012) discussed two problems of the standardization method, and it is therefore important that we address these in light of our Bayesian formulation. First, Franz and Loftus stated that the associated intervals are too small, as $SEM^{Norm}$ (where SEM is an abbreviation for standard error of the mean) underestimates the associated SEM produced by the Loftus and Masson (1994) method, $SEM^{L\&M}$, by a factor of $\sqrt{\frac{C-1}{C}}$. The rescaling proposed by Morey (2008) can be applied here, though we do not pursue this modification as it would alter the modified posterior probability of the resulting within-subject Bayesian HDI to a value above the nominal $1-\alpha$ level. From Theorem 3, the modified posterior probability of our proposed interval is guaranteed to be $1-\alpha$, and the length of the unadjusted interval will be smaller than that of the adjusted interval. It is also instructive to point out that the same term $\sqrt{\frac{C-1}{C}}$ also appears in equation (\ref{length_ratio}). 

The second problem of the standardization method discussed by Franz and Loftus (2012), is that the method can hide serious violations of the circularity assumption, that is, an assumption on the covariance matrix of the repeated measurements that the variance is constant and the covariance between any pair of measurements is also constant. It should therefore not be used as a tool to detect departures from circularity. We agree with this point, and suggest that the approach recommended by those authors (i.e., showing all pairwise differences between factor levels and computing the corresponding $SEM$ for each pair) can be employed as a simple diagnostic to check for the violation of circularity. Alternatively, various statistical packages (e.g., ezANOVA in the R package \emph{ez}) can be used directly to test the circularity assumption. 

To determine which of our proposed within-subject HDI's, either (\ref{new_HDI}) or (\ref{interval_HV}), to use for a given dataset, we recommend either simply inspecting the variability of the data at each level of the independent variable to determine whether homogeneity of variance is a reasonable assumption, or, more formally, comparing the underlying models (\ref{mixed_model}) and (\ref{mixed_H}) using Bayesian model selection procedures (e.g., Kass \& Raftery, 1995;  Wagenmakers, 2007; Masson, 2011; Rouder et al., 2012; Nathoo \& Masson, 2016). For example, the Bayes factor can be used to compare models (\ref{mixed_H}) and (\ref{mixed_model}), and its computation can be implemented using the BayesFactor package in R.

\section{Data Examples}
In this section we illustrate applications of our proposed computation of credible intervals for condition means in a repeated-measures design. For the first example, we consider the hypothetical data used by Loftus and Masson (1994) to demonstrate the application of the within-subject confidence interval they developed. The data consist of scores from 10 subjects, each tested under three conditions representing three different presentation durations (see Table 2 in Loftus \& Masson). In Table 1, we present the raw data and the means for each of the three conditions in their example. On the right side of the table, we present two versions of the 95\% confidence intervals, one being the standard confidence interval based on between-subject variability (assuming equal variance) and the other representing the within-subject CI defined by Loftus and Masson (\ref{LM}). Note that the within-subject CI is narrower than the between-subject CI because the within-subject version is computed with between-subject variability removed. We also present the 95\% within-subject highest density interval computed using (\ref{new_HDI}). This HDI reflects the credible values of the condition means, conditioned on the variability between subjects.

\singlespace
\begin{table}
\begin{tabular}[t]{rrrr}
   \hline
   \\[-1em]
   Subject & 1 sec & 2 sec & 5 sec\\
   \\[-1em]
   \hline
   1 & 10 & 13 & 13\\
   2 & 6 & 8 & 8\\
   3 & 11 & 14 & 14\\
   4 & 22 & 23 & 25\\
   5 & 16 & 18 & 20\\
   6 & 15 & 17 & 17\\
   7 & 1 & 1 & 4\\
   8 & 12 & 15 & 17\\
   7 & 1 & 1 & 4\\
   8 & 12 & 15 & 17\\
   9 & 9 & 12 & 12\\
  10 & 8 & 9 & 12\\
  \hline
  \\[-1em]
  Mean & 11.0 & 13.0 & 14.2\\
  \\[-1em]
  \hline
\end{tabular}
\quad
\begin{tabular}[t]{cc}
   \hline
   \\[-1em]
   Interval type & Interval width\\
   \\[-1em]
   \hline
   \\[-1em]
   95\% between- & \\
   subject CI  & $\pm$3.86\\[5pt]
   95\% within- & \\
   subject CI  & $\pm$0.52\\[5pt]
   95\% within- & \\
   subject HDI  & $\pm$0.42\\[5pt]
   \hline
\end{tabular}
\caption{\\Hypothetical Data from a Within-Subject Design and 95\% Confidence Intervals and 95\% Highest Density Interval.\\[-1em]}
\end{table}

\doublespace
To help elucidate the meaning of the within-subject HDIs that we have proposed, we make a comparison to a standard Bayesian HDI. One approach to computing standard HDIs for a within-subject design, would be to use the large-sample approximation to the posterior distribution presented by Nathoo and Masson (2016, p. 148). These HDIs constitute credible intervals for the condition means in a repeated-measures design. These intervals can be computed based on the observed condition mean, $M_{. j}$, and from components drawn from a standard ANOVA for a within-subjects design:
\begin{gather*}
   M_{. j} \pm (Z_{1-\alpha/2})SD_j\\
   \text{where,}\ SD_j = \frac{1}{N}\sqrt{\frac{SS_T-SS_C}{k}},
\end{gather*}
\noindent where $N$ = number of subjects, $k$ = number of conditions, $Z_{1-\alpha/2}$ is the $\alpha/2$ critical value of the standard normal distribution, and $SS_T$ and $SS_C$ are, respectively, the sum of squares total and sum of squares conditions from the repeated-measures ANOVA. 

For the example data set in Table 1, and using the equation above, the associated 95\% HDI has a width of $\pm$ 3.49. Notice that this credible interval is substantially larger than the within-subject HDI. As with the contrast between the two versions of CIs shown in Table 1, the reason for this difference is that the within-subject HDI is based on a modified posterior distribution that is conditioned on the between-subject variability, so that the length of the resulting interval does not take this variability into account. It is worth reiterating that this variability is not of scientific interest in within-subject designs. As a result, the within-subject HDI is not a credible interval in the usual sense; however, its modified posterior probability is $1-\alpha$.

Now consider a case in which the circularity assumption is violated by having unequal condition variances. Using the model described in (\ref{mixed_H}), we constructed a hypothetical set of response-time data for a within-subject design with three conditions and a sample of 48 subjects. In particular, the variance of one of the conditions was substantially larger than in the other two conditions. The descriptive statistics for these data are shown in Table 2. A clear indication of the violation of homogeneity of variance is the much larger variance in the third condition. 

Another more general indication of violation of the circularity assumption, is unequal variance of difference scores for pairs of conditions. To see this, difference scores can be computed for every subject on each pair of conditions, and the variance of difference scores for each condition pair can then be compared. If there is substantial inconsistency in these variances, either homogeneity of variance, or homogeneity of covariance (or both), has likely been violated. As shown in Table 2, the variance of difference scores for condition pairs involving the third condition is much higher than for the condition pair that excludes that condition.

\singlespace
\begin{table}
\begin{tabular}{rrrrrrrr}
   \hline
   \\[-1em]
   \multicolumn{1}{c}{} & \multicolumn{3}{c}{Condition} & \multicolumn{1}{c}{} & \multicolumn{3}{c}
   {$s^2$ of difference scores}\\
   \\[-1em]
   \multicolumn{1}{c}{} & \multicolumn{1}{c}{C1} & \multicolumn{1}{c}{C2} & \multicolumn{1}{c}{C3}
   & \multicolumn{1}{c}{} & \multicolumn{1}{c}{C1,C2} & \multicolumn{1}{c}{C1,C3} & \multicolumn{1}
   {c}{C2,C3}\\
    \cline{1-4}
    \cline{6-8} 
    \\[-0.75em]
    $M$ & 704 & 745 & 761 & & 411 & 21,113 & 22,182\\
    $s^2$ & 6,126 & 6,255 & 30,499 & & & & \\
     \\[-0.75em]
    \hline
    \\[-1em]
    \multicolumn{2}{l}{Source} & \multicolumn{1}{c}{$SS$} & \multicolumn{1}{c}{$df$} & \multicolumn{1}
    {c}{$MS$} & \multicolumn{1}{c}{$F$} & \multicolumn{2}{c}{}\\
    \\[-1em]
     \cline{1-6}
     \\[-0.75em]
     \multicolumn{2}{l}{Subjects} & \multicolumn{1}{r}{1,330,612} & \multicolumn{1}{r}{47} & 
     \multicolumn{4}{c}{}\\
     \multicolumn{2}{l}{Conditions} & \multicolumn{1}{r}{85,323} & \multicolumn{1}{r}{2} & 
     \multicolumn{1}{r}{42,662} & \multicolumn{1}{r}{5.86} & \multicolumn{2}{c}{}\\
     \multicolumn{2}{l}{SxC} & \multicolumn{1}{r}{684,728} & \multicolumn{1}{r}{94} & 
     \multicolumn{1}{r}{7,284} & \multicolumn{3}{c}{}\\
     \multicolumn{2}{l}{Total} & \multicolumn{1}{r}{2,100,663} & \multicolumn{1}{r}{143} & 
     \multicolumn{4}{c}{}\\
     \\[-0.75em]
     \hline
\end{tabular}
\hfill\
\caption{\\Descriptive Statistics, Variances of Difference Scores, and ANOVA for a Hypothetical Data Set Violating the Assumption of Homogeneity of Variance.\\[-1em]}
\end{table}

\doublespace
Our recommended approach in such situations, is to compute a Bayesian within-subject HDI based on the heteroscedastic mixed effects model using (\ref{interval_HV}). This procedure calls for each subject's score within a condition to be standardized relative to that subject's mean, as well as to the grand mean. As per (\ref{interval_HV}), each subject's raw score was standardized using the equation $Y_{ij}^{'}=Y_{ij}-M_{i.}+M$. The standard error of the mean for each condition was then computed based on these standardized scores, and served as the basis for the Bayesian within-subject HDI, as per (\ref{interval_HV}). As can be seen in Figure 1, the resulting HDIs varied in size across the three conditions. For comparison, we also show in this figure, the within-subject HDI computed under the assumption of equal variance based on (\ref{new_HDI}). A key difference in these two approaches to constructing within-subject credible intervals, is that when heteroscedasticity is taken into account, the consequence of the high variability of scores in one condition is assigned \emph{specifically to that condition}, instead of being spread across all conditions as happens when homogeneity is assumed.

\begin{figure}[h]
  \centering
  \includegraphics{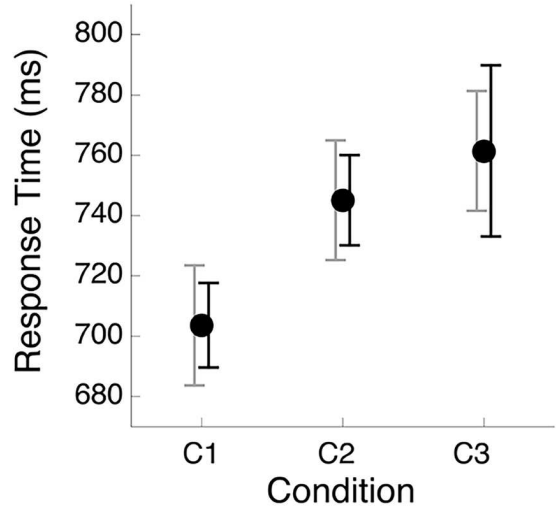}
  \caption{Means and Bayesian within-subject highest density intervals (HDIs) for a hypothetical set of response-
  time data for three conditions of a within-subject design. One set of within-subject HDIs was computed by 
  assuming the circularity assumption holds (grey bars) and the other set was based on the
  assumption that circularity was violated by heteroscedasticity (black bars).}
\end{figure}

\section{Discussion}

We have developed a Bayesian approach to within-subject interval estimation for repeated-measures designs. Our primary contributions are fourfold. First, we have defined the notion of a within-subject Bayesian HDI based on a conditional posterior distribution. This distribution is based on conditioning and applying the maximum likelihood estimates of the subject-specific random effects. Second, we have shown that the original within-subject CI can be viewed as a within-subject Bayesian HDI corresponding to a particular improper prior, and have further demonstrated that this HDI has a modified posterior probability of $1-\alpha$. This contribution sheds new light on how to interpret the original within-subject interval. Third, we have proposed a new within-subject Bayesian HDI based on a standard noninformative prior, and argued that the new interval is a better within-subject interval estimate. Fourth, we have considered the case of heteroscedastic data, and have derived the standardization method as a Bayesian within-subject HDI for the heteroscedastic case. The new formulation of this method allows for a new interpretation in terms of its conditional posterior probability. 

Although we have considered only single-factor mixed effects models for Gaussian data in this paper, several extensions will be useful to consider for future work in the new paradigm of within-subject Bayesian inference. First, it would be useful to extend our approach to multi-factor within-subject designs and designs having within-subject and between-subject factors. 

Second, the underlying model can be extended to allow for outliers or skewed data. Along these lines, Kennedy, Navarro, Perfors, and Briggs (2017) have recently considered the robustness of Bayesian credible intervals to violations of standard modelling assumptions. In line with expectations, these authors find that models based on the Gaussian distribution, such as the mixed effects models that we consider here, are not robust to contamination. This problem is taken care of in many applications of our method to psychological experiments, as outlier removal techniques are used as a pre-processing step (i.e., data cleaning). As a result, the data submitted to the computation of intervals have had outliers removed prior to averaging across observations. Nevertheless, developing the within-subject Bayesian HDI with models based on distributions having more flexible tail behavior could be a useful extension, and the incorporation of flexible parametric distributions is one such approach for future development. For example, skew-elliptical distributions are a convenient possibility since the Gaussian-location-scale-mixture representation of skew-elliptical distributions (see, e.g., Branco \& Dey, 2001; Nathoo \& Ghosh, 2012) typically allows for a straightforward computational implementation within standard MCMC algorithms. 

Third, for within-subject designs with a binary response, logistic mixed models are often used. For example, Song, Nathoo, and Masson (2017) discuss Bayesian inference in this case and provide software for Bayesian hypothesis testing and estimation in logistic and probit mixed models for within-subject designs. Within this context it may be useful to consider within-subject HDI's for the experimental effects in logistic mixed models.  

With regard to the simple estimates of the random effects used here, a referee has correctly pointed out that these estimates have no shrinkage. Our choice of plug-in estimator is based on a desire for simple closed forms that can be easily used by practitioners. Future work (e.g. investigating extensions of the within-subject interval to mixed logistic models) will consider alternative potentially improved shrinkage estimation for the subject-specific random effects and will investigate the impact of such estimators on the performance of the within-subject HDI.

Finally, we offer some observations on the appropriate interpretation of the within-subject HDI that we have introduced. We reiterate that there is a notion of '\emph{false certainty}' associated with the within-subject interval estimate, since a legitimate source of uncertainty has been deliberately removed. One simple way to deal with this concern is to report, in addition to the $1-\alpha$ within-subject HDI, the unconditional posterior probability $Pr(L_{j} \le \mu_{j} \le U_{j}|\boldsymbol{Y})$ of that same interval. The practical justification for the construction of the within-subject HDl, is that it excludes the component of variability that is not of scientific interest in within-subject designs. It can thus be used as a graphical tool capable of revealing a consistent pattern of within-subject effects (that otherwise would likely be hidden), while also depicting the source of posterior variability that is of scientific interest in these designs. 

In contrast to the case of within-subject CIs developed by Loftus and Masson (1994), where the CIs stand in a simple relationship to the outcome of the corresponding significance test (i.e., if two means are separated by at least $\sqrt{2}$ times the size of one side of the CI, then the means are significantly different from each other), no such simple relationship has been established between within-subject HDIs and measures of the strength of evidence for effects (such as Bayes factors). At this stage of development, these HDIs can be used to convey a general impression of the plausibility of the observed separation between means. Even this information is much more useful than providing potentially misleading error bars that are heavily influenced by between-subject variability that bears no relevance to the question being examined, or worse, providing no error bars at all.

\section*{Competing Interests}
\noindent The authors declare that they have no competing interests.

\section*{Appendix}

\noindent {\bf Proof of Theorem 1:}\\
We let $\boldsymbol{\mu} = (\mu_{1},\dots,\mu_{C})^{T}$ and derive the form for the conditional distribution  [$\boldsymbol{\mu}\mid\boldsymbol{Y},b_1,\dots\,,b_N$] and then plug-in the estimates  $\hat{b}_{i} = M_{i .} - M$. In this case the prior for the model parameters takes the form $\pi(\boldsymbol{\mu},\sigma_{\epsilon}^{2})\propto\frac{1}{\sigma_{\epsilon}^{2}}$, and  the conditional posterior takes the form
\begin{gather*}
   p(\boldsymbol{\mu},\sigma_{\epsilon}^{2}\mid \boldsymbol{Y}, \boldsymbol{b})\propto
   \biggl[\prod_{i=1}^{N}\prod_{j=1}^{C}p(Y_{ij}\mid \mu_j, b_i, \sigma_{\epsilon}^{2})\biggr]
   \pi(\boldsymbol{\mu},\sigma_{\epsilon}^{2})\\
   \propto \biggl[\prod_{i=1}^{N}\prod_{j=1}^{C}\text{N}(y_{ij};\mu_j + b_i, \sigma_{\epsilon}^{2})\biggr]\frac{1}
   {\sigma_{\epsilon}^{2}}\\
   \propto \biggl[\prod_{i=1}^{N}\prod_{j=1}^{C}\frac{1}{\sigma_{\epsilon}}\exp{\Bigl\{-\frac{1}
   {2\sigma_{\epsilon}^{2}}(y_{ij}-\mu_j-b_i)^2\Bigr\}}\biggr]\frac{1}{\sigma_{\epsilon}^{2}}\\
   \propto \biggl[\frac{1}{\sigma_{\epsilon}^{2}}\biggr]^{\frac{NC}{2}+1}\exp{\biggl\{-\frac{1}
   {2\sigma_{\epsilon}^{2}}\sum_{i=1}^{N}\sum_{j=1}^{C}(y_{ij}-\mu_j-b_i)^2\biggr\}}
\end{gather*}
\noindent then
\begin{equation*}
   p(\boldsymbol{\mu}\mid \boldsymbol{Y}, \boldsymbol{b})\propto
   \underbrace{\int_0^\infty \biggl[\frac{1}{\sigma_{\epsilon}^{2}}\biggr]^{\frac{NC}{2}+1}\exp{\biggl\{-\frac{1}
   {2\sigma_{\epsilon}^{2}}\sum_{i=1}^{N}\sum_{j=1}^{C}(Y_{ij}-\mu_j-b_i)^2\biggr\}}d
   \sigma_{\epsilon}^{2}}_{\text{(inverse-gamma integral)}}
\end{equation*}
\begin{gather*}
   \propto \biggl(\sum_{i=1}^{N}\sum_{j=1}^{C}(Y_{ij}-\mu_j-b_i)^2\biggr)^{-NC/2}\\
   \propto \biggl(\sum_{j=1}^{C}\sum_{i=1}^{N}(Y_{ij}-\mu_j-b_i)^2\biggr)^{-\bigl(C(N-1)+C\bigr)/2}
\end{gather*}
\noindent and after some algebra we get
\begin{equation*}
   \propto \biggl(1+ \frac{N(N-1)CO^{-1}}{(N-1)C}\bigl(\boldsymbol{\mu} - \frac{1}{N}\boldsymbol{Yb}
   ^{(\tau)}\bigr)^{T}\bigl(\boldsymbol{\mu} - \frac{1}{N}\boldsymbol{Yb}^{(\tau)}\bigr)\biggr)^
   {-\bigl(C(N-1)+C\bigr)/2}
\end{equation*}\
   where $\boldsymbol{Yb}^{(\tau)} = (Yb_1^{(\tau)},\dots\,,Yb_C^{(\tau)})^{T},\;\; 
   Yb_j^{(\tau)} = \sum_{i=1}^N(Y_{ij}-b_i)$
   and $O=\sum_{j=1}^C\sum_{i=1}^N(Y_{ij}-b_{i})^2 - \frac{1}{N}\boldsymbol{Yb}^{(\tau)^T}
   \boldsymbol{Yb}^{(\tau)}$.

From this form we see that $\boldsymbol{\mu}\mid\boldsymbol{Y},\boldsymbol{b} \sim MVt_{\nu}(\boldsymbol{\mu}_{\mu}, \boldsymbol{\Sigma})$, the multivariate t-distribution,
where
\begin{gather*}
   \nu = C(N-1), \boldsymbol{\mu}_{\mu} = \frac{1}{N}\boldsymbol{Yb}^{(\tau)}, \boldsymbol{\Sigma} = \frac{1}{N(N-1)C}\cdot O\cdot\boldsymbol{I}.
\end{gather*}
The within-subject HDI's are then based on $p(\boldsymbol{\mu}\mid\boldsymbol{Y},\hat{b})$ where $\hat{b}_i = M_{i .}-M$. Plugging in $\hat{b}_i$ for $b_{i}$  we get $\boldsymbol{\hat{\mu}}_{\mu} = (M_{.1},\dots\,,M_{.C})^{T}$, the condition means, and after some more algebra, one can show that
\begin{equation*}
   \hat{O} = SS_{S\times C} = \sum_{j=1}^C\sum_{i=1}^NY_{ij}^2 - C\sum_{i=1}^NM_{i.}^2 - N\sum_{j=1}
   ^CM_{. j}^2 + CNM^2.
\end{equation*}
We then have that
\begin{center}
   $\boldsymbol{\mu}\mid\boldsymbol{Y},\hat{b} \;\; \sim \;\;MVt_{C(N-1)}\bigl((M_{. j}),
   \frac{SS_{S\times C}}{N(N-1)C}\cdot\boldsymbol{I}\bigr)$.
\end{center}
From this it follows that the marginal posterior distribution is
\begin{gather*}
   \mu_j\mid\boldsymbol{Y},\boldsymbol{\hat{b}}\sim t_{C(N-1)}\bigl(M_{. j},\frac{SS_{S\times C}}
   {N(N-1)C}\bigr)
\end{gather*}
as claimed. $\blacksquare$

\bigskip

\noindent {\bf Proof of Corollary 1:}\\
From Theorem 1 we can write
\begin{gather*}
   \frac{\mu_j - M_{. j}}{\sqrt{\frac{SS_{S\times C}}{N(N-1)C}}}\;\;\big|\;\;\boldsymbol{Y},\boldsymbol{\hat{b}}
   \;\;\sim\;\;t_{C(N-1)}(0,1)
\end{gather*}
which then leads to the following equation
\begin{gather*}
Pr\Biggl(-[\textup{criterion}\ t_{C(N-1)}]_{\frac{\alpha}{2}}\;\;\leq\;\;\frac{\mu_j - M_{. j}}
   {\sqrt{\frac{SS_{S\times C}}{N(N-1)C}}}\;\;\leq\;\;[\textup{criterion}\ t_{C(N-1)}]_{\frac{\alpha}{2}}\;\;\big|
   \;\;\boldsymbol{Y},\boldsymbol{\hat{b}}\Biggr)
   = 1 - \alpha
\end{gather*}
which after isolating for $\mu_{j}$ yields the following $1-\alpha$ conditional posterior interval  for $\mu_j$ 
\begin{equation*}
    M_{. j}\pm\sqrt{\frac{SS_{S\times C}}{N(N-1)C}}\;\;[\textup{criterion}\ t_{C(N-1)}]_{\frac{\alpha}{2}}.
\end{equation*}
Since the marginal posterior distribution in Theorem 1 is unimodal and symmetric about its mode $M_{. j}$, it follows that this interval is the within-subject $1-\alpha$ HDI for $\mu_{j}$. $\blacksquare$

\bigskip

\noindent {\bf Proof of Theorem 2:}\\
The proof follows immediately after replacing the prior $\pi(\boldsymbol{\mu},\sigma_{\epsilon}^{2})\propto\frac{1}{\sigma_{\epsilon}^{2}}$ with the prior $\pi(\boldsymbol{\mu}, \sigma_{\epsilon}^2)\propto\Bigl[\frac{1}{\sigma_{\epsilon}^2}\Bigr] ^{\frac{-N+3}{2}}$ and following exactly the steps in the proofs of Theorem 1 and Corollary 1.  $\blacksquare$

\bigskip

\noindent {\bf Proof of Theorem 3:}\\
The prior distribution is taken as $ \pi(\boldsymbol{\mu}, \sigma_1^2,\dots,\sigma_C^2)\propto\prod_{j=1}^C\sigma_j^{-2}$ and under the heteroscedastic mixed effects model we have\\
\begin{gather*}
   p(\boldsymbol{\mu},\boldsymbol{\sigma^2}\mid\boldsymbol{Y},\boldsymbol{\hat{b}})\propto 
   p(\boldsymbol{Y}\mid\boldsymbol{\mu},\boldsymbol{\sigma^2},\boldsymbol{\hat{b}})
   \pi(\boldsymbol{\mu},\boldsymbol{\sigma^2})\\
   \propto\biggl[\prod_{i=1}^N\prod_{j=1}^Cp(Y_{ij}\mid\mu_j, \sigma_j^2, \hat{b}_i)\biggr]
   \times\biggl[\prod_{j=1}^C\sigma_j^{-2}\biggr]\\
   \propto\prod_{j=1}^C\biggl[\prod_{i=1}^Np(Y_{ij}\mid\mu_j, \sigma_j^2, \hat{b}_i)\biggr]
   \sigma_j^{-2}
   \propto\prod_{j=1}^C\biggl[\prod_{i=1}^N(\sigma_j^{-2})^{\frac{1}{2}}\exp{\Bigl\{-\frac{1}
   {2\sigma_j^2}(Y_{ij}-\mu_j-\hat{b}_i)^2}\Bigr\}\biggr]\sigma_j^{-2}\\
   \propto\prod_{j=1}^C(\sigma_j^{-2})^{\frac{N}{2}+1}\exp{\Bigl\{-\frac{1}{2\sigma_j^2}
   \sum_{i=1}^N(Y_{ij}-\mu_j-\hat{b}_i)^2}\Bigr\}
\end{gather*}
\begin{gather*}
   \rightarrow p(\boldsymbol{\mu}\mid\boldsymbol{Y},\boldsymbol{\hat{b}})\propto\int_0^\infty
   \dots\int_0^\infty\prod_{j=1}^C(\sigma_j^{-2})^{\frac{N}{2}+1}\exp{\Bigl\{-\frac{1}{2\sigma_j^2}
   \sum_{i=1}^N(Y_{ij}-\mu_j-\hat{b}_i)^2}\Bigr\}d\sigma_1^2\dots d\sigma_C^2\\
   \propto\prod_{j=1}^C\underbrace{\int_0^\infty(\sigma_j^{-2})^{\frac{N}{2}+1}\exp{\Bigl\{-\frac{1}
   {2\sigma_j^2}\sum_{i=1}^N(Y_{ij}-\mu_j-\hat{b}_i)^2}\Bigr\}d\sigma_j^2}_{\text{(inverse-gamma integral)}}\\
   \propto\prod_{j=1}^C\biggl(\sum_{i=1}^N(Y_{ij}-\mu_j-\hat{b}_i)^2\biggr)^{-\frac{N}{2}}\\
   \propto\prod_{j=1}^C\Biggl(1+\frac{N\nu}{\nu}\biggl(\sum_{i=1}^N(Y_{ij}-M_{i .}+M)^2-
   NM_{. j}^2\biggr)^{-1}(\mu_j-M_{. j})^2\Biggr)^{-\frac{(\nu+1)}{2}}\\
   (\text{where}\ \nu=N-1)\\
   \propto\prod_{j=1}^C\underbrace{\Biggl(1+\frac{N(N-1)}{(N-1)}\biggl(\sum_{i=1}^N(Y_{ij}-
   M_{i .}+M)^2-NM_{. j}^2\biggr)^{-1}(\mu_j-M_{. j})^2\Biggr)^{-\frac{(N-1+1)}{2}}}_{\text{Proportional to the density of a }t_{N-1}(M_{. j},\ \omega_j^2) \text{ distribution}}
\end{gather*}
\begin{center}
   where\;\;$\omega_j^2=\frac{1}{N(N-1)}\sum_{i=1}^N(Y_{ij}-M_{i .}+M)^2-NM_{. j}^2$.\\[10pt]
\end{center}
Thus we have $\mu_{j} \mid \boldsymbol{Y},\boldsymbol{\hat{b}} \stackrel{ind}{\sim} t_{N-1}(M_{. j},\ \omega_j^2), \, j=1, \dots, C$.
It then follows that
\begin{center}
\;\;$\frac{\mu_j-M_{. j}}{\omega_j}\mid\boldsymbol{Y},\boldsymbol{\hat{b}}\;\;\sim\;\; t_{N-1}(0,1)$\\
\end{center}
so that the following equation holds
\begin{gather*}
   Pr\biggl(-[\text{criterion}\ t_{N-1}]_{\frac{\alpha}{2}}\leq\frac{\mu_j-M_{. j}}{\omega_j}\leq
   [\text{criterion}\ t_{N-1}]_{\frac{\alpha}{2}}\mid\boldsymbol{Y},\boldsymbol{\hat{b}}\biggr)=1-\alpha.
\end{gather*}
After isolating for $\mu_{j}$, a $1-\alpha$ posterior interval for $\mu_{j}$ is obtained as
\begin{gather*}
M_{. j}\pm\omega_j[\text{criterion}\ t_{N-1}]_{\frac{\alpha}{2}}\\
   \rightarrow M_{. j}\pm\biggl[\frac{1}{N(N-1)}\sum_{i=1}^N(Y_{ij}-M_{i .}+M)^2-NM_{. j}^2\biggr]^{\frac{1}{2}}
   [\text{criterion}\ t_{N-1}]_{\frac{\alpha}{2}}\\
   \rightarrow M_{. j}\pm\biggl[\frac{1}{N(N-1)}\sum_{i=1}^N(Y_{ij}^{'}-M_{. j})^2\biggr]^{\frac{1}{2}}
   [\text{criterion}\ t_{N-1}]_{\frac{\alpha}{2}}\\
   \text{where}\ Y_{ij}^{'}=Y_{ij}-M_{i .}+M
\end{gather*}
so that,
\begin{gather*}
    M_{. j}\pm SEM_j^{Norm}[\text{criterion}\ t_{N-1}]_{\frac{\alpha}{2}}\\
   \text{where}\;\;SEM_j^{Norm}=\sqrt{\frac{1}{N(N-1)}\sum_{i=1}^N(Y_{ij}^{'}-M_{. j})^2}\\
   Y_{ij}^{'}=Y_{ij}-M_{i .}+M,
   \text{ is a}\ 100(1-\alpha)\%\ \text{posterior interval.} 
\end{gather*}
Since the marginal posterior distribution is unimodal and symmetric about its mode $M_{. j}$, it follows that this interval is the within-subject $1-\alpha$ HDI for $\mu_{j}$ in the heteroscedastic case. $\blacksquare$

\section*{References}
\nocite{*}
\bibliography{NathooEtAl2017_JMP}
\bibliographystyle{apalike}

\end{document}